\documentclass[reprint,superscriptaddress,amsmath,amssymb,aps,prl]{revtex4-2}
\usepackage{physics}
\usepackage{amsmath, amsfonts}
\usepackage{float}
\usepackage{graphicx} 
\usepackage{dcolumn} 
\usepackage{bm} 
\usepackage{xcolor}

\begin{document}

\title{Quantum Geometric Origin of Strain-Induced Ferroelectric Phase Transitions}

\author{Jiaming Hu}
 \affiliation{Center for Quantum Matter, School of Physics, Zhejiang University, Hangzhou 310058, China.}

\author{Ziye Zhu}
\affiliation{School of Engineering, Westlake University, Hangzhou 310030, China.}

\author{Yubo Yuan}
\affiliation{School of Engineering, Westlake University, Hangzhou 310030, China.}

\author{Wenbin Li}
\affiliation{School of Engineering, Westlake University, Hangzhou 310030, China.}
\affiliation{Zhejiang Key Laboratory of 3D Micro/Nano Fabrication and Characterization, Westlake University, Hangzhou 310030, China}

\author{Hua Wang}
\email{daodaohw@zju.edu.cn}
\affiliation{Center for Quantum Matter, School of Physics, Zhejiang University, Hangzhou 310058, China.}

\author{Kai Chang}
\email{kchang@zju.edu.cn}
\affiliation{Center for Quantum Matter, School of Physics, Zhejiang University, Hangzhou 310058, China.}

\date{\today}

\begin{abstract}
    Strain-regulated ferroelectric (FE) materials have long attracted significant attention due to their diverse applications. While soft-phonon theory and the (pseudo) Jahn-Teller effect have achieved considerable success in providing phenomenological descriptions and general understanding, the detailed connection between these perspectives and their microscopic dependence on strain regulation remains unclear. Here, under the framework of density-functional perturbation theory (DFPT), we demonstrate that the Berry curvature of electron-phonon coupling (EPC) plays a pivotal role in the interatomic force matrix (IFM). A subsequent model analysis shows that external strain can reverse the polarity of the EPC Berry curvature in (quasi)-degenerate electronic subsystems through band inversion, thereby directly leading to phonon softening. The general theory is then applied to the BiOCl monolayer as a benchmark, which offers an accurate description of the density functional theory (DFT) calculations. This mechanism is further observed across a broad range of materials through ab initio calculations, providing an insightful perspective on EPC quantum geometry in lattice dynamics and FE phase transitions.
\end{abstract}

\maketitle


\textit{Introduction.}---Ferroelectric (FE) materials have attracted widespread interest for decades due to their promising applications across various fields~\cite{martin2016thin,wei2022progress,han2022ferroelectric,setter2006ferroelectric,shkuratov2022review,blazquez2018biological}. In particular, strain-regulated FE phase transitions have garnered significant attention and have been observed in a wide range of materials~\cite{zhu2024highly,pesquera2020beyond,xu2020strain,behara2022ferroelectric,schlom2007strain,ederer2005effect}, making it crucial to understand their microscopic origin. To explain these FE transitions, the soft-phonon theory~\cite{shirane_et_al_1967,pytte_1972,rowley_et_al_2018,misra_et_al_2005,PhysRevLett.22.1251,PhysRevLett.107.266401,PhysRevLett.40.465,PhysRevB.13.271} has been widely adopted, which states that FE phase transitions occur when the energy of certain optical phonon modes at the $\Gamma$ point is reduced to zero. Although this approach offers a fundamental phenomenological description, it does not elucidate the microscopic mechanism driving these transitions. On the other hand, the (pseudo) Jahn-Teller effect is recognized as the central mechanism for FE instability~\cite{bersuker2020jahn,bersuker2013jahn}. 
This effect attributes spontaneous lattice distortion to the electron-phonon coupling (EPC) of degenerate energy bands~\cite{kristoffel1973electron}. However, the requirement for band crossing limits its applicability to materials with gap closures. The application of the (pseudo) Jahn-Teller effect to large-gap FE materials, along with its strain-regulation behavior and direct correlation with soft phonons, remains ambiguous. To address this challenge, it is promising to draw upon the insights provided by quantum geometry. The geometric properties of quantum states are increasingly recognized as critical for understanding various condensed-matter phenomena~\cite{ahn2022riemannian,xiao2010berry,nagaosa2010anomalous,sundaram1999wave,resta2011insulating,PhysRevLett.49.405,sodemann2015quantum,citro2023thouless,wang2022generalized,PhysRevX.10.041041,gu2023discovery,morimoto2016topological,morimoto2016semiclassical,bhalla2022resonant}. Recent studies have highlighted their importance in the phonon Hall effect~\cite{im2022ferroelectricity}, chiral phonons~\cite{saparov2022lattice}, EPC~\cite{yu2024non,hu2024phononmediatednonlinearopticalresponses}, and FE polarization~\cite{onoda2004topological,liu2016strain,citro2023thouless,liu2016strain}, motivating further investigation into their role in FE phase instability. 

In this Letter, we unveil the quantum geometric origin of strain-induced FE instability by: (i) elucidating the contribution of EPC Berry curvature to the inter-atomic force matrix (IFM) within the framework of density-functional perturbation theory (DFPT), (ii) identifying its polarity reversal as the source of soft $\Gamma$ phonon using a generalized model, and (iii) applying these insights to explain the geometric origin of strain-tunable FE instability in 2D-layer BiOX (X = Cl, Br, I). The validity of our explanation is further corroborated by comparisons with \textit{ab initio} calculations and extended to a broad range of materials.

\textit{Inter-atomic force matrix and quantum geometry.}---A crystal lattice is defined by the positional vectors of $N_c$ ions, $\{\bm{R}_i \mid i=1,2,\ldots,N_c\}$. The equilibrium condition requires that the Hellmann-Feynman force $L^a_i$ acting on the $i$-th ion in the $a = x, y, z$ direction vanishes. The IFM, defined as the Jacobian matrix of the ionic forces, must be positive-definite for dynamic stability. Within the DFPT framework, the IFM element corresponding to the displacements of the $i$-th and $j$-th ions in the $a$ and $b$ directions is expressed as~\cite{RevModPhys.73.515}: 
\begin{equation}\label{eq:raw_Cs_main}
    \begin{aligned}
        C_{ij}^{ab} = \frac{\partial L^a_i}{\partial R^b_j} &= \int d\bm{r}
        \frac{\partial n(\bm{r})}{\partial R_i^a}\frac{\partial V_{\rm ext}(\bm{r})}{\partial R_j^b}
        + \int d\bm{r}n(\bm{r})\frac{\partial^2 V_{\rm ext}(\bm{r})}{\partial R_i^a \partial R_j^b}
        \\
        &+ \int d\bm{r}\frac{\partial^2 K(\bm{r})}{\partial R^a_i \partial R^b_j}, 
    \end{aligned}
\end{equation}
where $n(\bm{r})$ is the electronic density at position $\bm{r}$, $R_{i}^{a}$ denotes the $a$-component of the positional vector of the $i$-th ion, $V_{\rm ext}(\bm{r})$ is the ionic potential field referring to electron-ion interaction, while $K(\bm{r})$ is the pure-ionic contribution. The second and third terms in Eq.~\ref{eq:raw_Cs_main}, denoted as $C^{ab}_{ij}|_{2,3}$, represent the classical response of the ionic system, which are basically not supposed to have polarity-reversal variation induced by regulations that conserve the mirror symmetry of crystal lattice along $a,b$ directions (details in Sec.II~\cite{SI_info}). In contrast, the first term, denoted as $C^{ab}_{ij}|_1$, captures the quantum mechanical contribution from perturbed electronic system. This kind of response usually relates to the phase aspect of electronic states, and can yield anomalous behaviors particularly for crossing bands~\cite{PhysRevLett.49.405,sodemann2015quantum,gao2014field,citro2023thouless,xiao2010berry}. Additionally, $C^{ab}_{ij}|_1$ may exhibit asymmetry under the exchange of spatial directions $a$ and $b$, which is responsible for the electrically induced Dzyaloshinskii-Moriya interaction (eDMI) effect~\cite{chen2022microscopic}, suggesting a unique contribution to lattice dynamics. Furthermore, the electronic system's response to external perturbations is closely tied to the quantum geometry of the corresponding parametric space. This aspect, to the best of our knowledge, has not been thoroughly investigated in the field of lattice dynamics, indicating a potential oversight to quantum geometric effect. 

\begin{figure}[htp]
    \centering
    \includegraphics[width=0.48\textwidth]{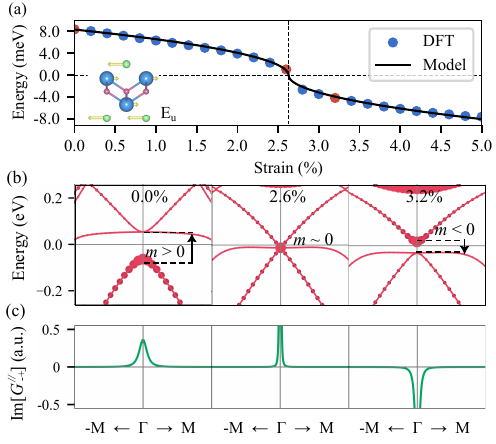}
    \caption{\label{fig:structure_frequency} \textbf{Strain-induced FE instability in BiOCl. } \textbf{(a)} Strain-regulated $E_u$ soft-mode phonon energy, with negative value referring to the imaginary phonon. The critical strain is around 2.6\%. \textbf{(b)} DFT-calculated energy bands of FE-inducing states, under different strain conditions highlighted as red dots in (a). The zero energy is 1.5${\sim}$1.6 eV below the Fermi level. The size of the filled circles represents the relative weight of $p_z$ orbit. $m$ is the energy difference between $p_z$ and $p_{xy}$ orbits. \textbf{(c)} Corresponding EPC Berry curvature ${\rm Im}[G^{\parallel}_{-+}]$ calculated by TB model given in Sec.~V~C of supplemental material~\cite{SI_info}. }
\end{figure}


Building upon the previous discussion, we focus on $C^{ab}_{ij}|_1$ to explore its quantum geometric properties. Our subsequent discussion will center on phonon displacements that preserve the translational symmetry of the static crystal lattice (i.e., those associated with $\Gamma$ phonons), as they are directly linked to FE instability. We begin by reformulating the electronic density as $
    n(\bm{r}) = \int [d\bm{k}] \sum_m f_m \langle m | \bm{r} \rangle \langle \bm{r} | m \rangle$,
which is contributed by all the eigenstates $|m\rangle$ weighted by their Fermi-Dirac occupations $f_m$. Here, $m$ labels both the band and wave vector $\bm{k}$, with $\int [d\bm{k}] = (2\pi)^{-d} \int d^d\bm{k}$ denoting the integral over the first Brillouin zone (BZ) in $d$ dimensions. With details presented in Sec.~I~\cite{SI_info}, we obtain: 
\begin{equation}\label{eq:C_ij_ab_braket}
    \begin{aligned}
        C_{ij}^{ab}|_1 
        &=
        \int [d\bm{k}]
        \sum_{m,n}
        2f_m \operatorname{Re}[{\langle} \partial^a_i m | n {\rangle} {\langle} n | \partial^b_j \hat{V}_{\rm ext} | m {\rangle}]
        \\
        &=
        \int [d\bm{k}]
        \sum_{m,n}
        f_{mn} \operatorname{Re}[{\langle} \partial^a_i m | n {\rangle} {\langle} n |\partial^b_j \hat{V}_{\rm ext} | m {\rangle}], 
    \end{aligned}
\end{equation}
where we use the shorthand $\partial/\partial R_i^a = \partial_i^a$ and the ionic-potential operator is defined as $\hat{V}_{\rm ext} = \int d\bm{r} |\bm{r}\rangle V_{\rm ext}(\bm{r}) \langle \bm{r} |$. The final line of Eq.~\ref{eq:C_ij_ab_braket} indicates that, in gapped systems, only transitions between valence and conduction states contribute to $C_{ij}^{ab}|_1$, highlighting its quantum excitation nature.

We further express $V_{\rm ext}(\bm{r})$ as a sum of single-ion potentials $V_{\rm ext}(\bm{r}) = {\sum\limits}^{N_c}_lv_l\big{(}|\bm{r}-\bm{R}_l|\big{)}$~\cite{PhysRevB.82.165111}, where $v_l\big{(}|\bm{r}-\bm{R}_l|\big{)}$ is the contribution of the $l$-th ion only depending on $|\bm{r}-\bm{R}_l|$ due to its spherical symmetry. As discussed in Sec.~I~\cite{SI_info}, in cases where the wavefunction overlap of two states $m,n$ is localized near the position $\bm{\zeta}_{mn}$ with a spread $\eta_{mn}$ (e.g. in the Wannier representation~\cite{PhysRevB.65.035109,giustino2007electron}), we approximate ${\partial^b_j}V_{\rm ext}(\bm{r})$ up to the first order of $\bm{r}-\bm{\zeta}_{mn}$ in the vicinity of $|\bm{r}-\bm{\zeta}_{mn}|<\eta_{mn}$, so that its inter-band matrix element becomes: 
\begin{equation}\label{eq:pbj_V_ext_nm_rb_main}
    \begin{aligned}
        {\langle}n|{\partial^b_j}\hat{V}_{\rm ext}|m{\rangle} = \bm{\gamma}^b_{j,nm}{\cdot}\bm{r}_{nm}+o(|\frac{\bm{r}-\bm{\zeta}_{mn}}{\bm{\zeta}_{mn}-\bm{R}_j}|^2), 
    \end{aligned}
\end{equation}
which is the product of two components: (i) The effective ionic force gradient in $b$-direction $\bm{\gamma}^b_{j,nm} = -{\partial_{r^b}}{\nabla_{\bm{r}}}v_j|_{\bm{r}=\bm{\zeta}_{mn}-\bm{R}_j}$ ($r^b$ is the $b$-component of positional vector $\bm{r}$). As an illustrative example presented in Fig.~S1~\cite{SI_info}, for localized valence electrons with $|\bm{\zeta}_{mn}-\bm{R}_j|{\gg}\eta_{mn}$, the classical electron-ion interaction is always a smooth function with great linearity near $\bm{\zeta}_{mn}$, making $\bm{\gamma}^b_{j,mn}$ a well-defined leading term; (ii) The inter-band matrix element of positional operator $\bm{r}_{nm}$, which refers to the $\bm{k}$-space Berry connection in crystalline solids as~\cite{resta2011insulating}: 
\begin{equation}\label{eq:r_b_mn_main}
    \bm{r}_{nm} = i{\langle}n|{\nabla_{\bm{k}}}m{\rangle}. 
\end{equation}Consequently, based on Eqs.~\ref{eq:pbj_V_ext_nm_rb_main} and~\ref{eq:r_b_mn_main}, we can reformulate Eq.~\ref{eq:C_ij_ab_braket} into the following form:
\begin{equation}\label{eq:C_ij_ab_1_geo_main}
    \begin{aligned}
        C_{ij}^{ab}|_1 
        &= 2 \int [d\bm{k}] \sum_{mn} f_m \bm{\gamma}^b_{j,nm} \cdot {\rm Im} \left[ \langle \partial^a_i m | n \rangle \langle n | \nabla_{\bm{k}} m \rangle \right].
    \end{aligned}
\end{equation}
Transforming this expression to the eigenspace of the $s$-mode phonon using the mass-normalized, dimensionless eigenvector $\bm{U}^s = (\{U^s_{i,a}\})^{\bm{T}}$, we obtain our central result:
\begin{equation}\label{eq:C_ij_ab_1_geo_s_main}
    \begin{aligned}
        C^s|_1 
        &= \sum_{ij,ab} U^s_{i,a} C_{ij}^{ab}|_1 U^s_{j,b} \\
        &= 2 \int [d\bm{k}] \sum_{mn} f_m \bm{\gamma}^s_{nm} \cdot {\rm Im}[\bm{G}^s_{mn}],
    \end{aligned}
\end{equation}
where $\bm{\gamma}^s_{nm} = \sum_{j,b} U^s_{j,b} \bm{\gamma}^b_{j,nm}$ is the effective force gradient of the $s$-mode phonon vibration, and $\bm{G}^s_{mn}$ is the \textit{EPC quantum geometric tensor} defined as:
\begin{equation}
    \bm{G}^s_{mn} = \langle \partial_s m | n \rangle \langle n | \nabla_{\bm{k}} m \rangle,
\end{equation}
with ${\partial_s} = \sum_{i,a} U^s_{i,a} \partial^a_i$ being the simplified notation for the partial derivative with respect to the $s$-mode phonon vibration. The tensor $\bm{G}^s_{mn}$ serves as the quantum geometric tensor in the hybrid Hilbert space parameterized by the $s$-mode phonon displacement and the electronic wavevector $\bm{k}$. It can be experimentally detected through the phonon-mediated optical responses~\cite{hu2024phononmediatednonlinearopticalresponses} since it has a fundamental relation with the $s$-mode EPC matrix element $g^s_{mn}$ and velocity matrix element $\bm{v}_{nm}$ as~\cite{hu2024phononmediatednonlinearopticalresponses}:
\begin{equation}
    \bm{G}^s_{mn} = -\hbar \frac{g^s_{mn} \bm{v}_{nm}}{\bar{Q}_s \epsilon^2_{mn}},
\end{equation}
where $\bar{Q}_s = \sqrt{\hbar / 2M \omega_s}$ is the phonon displacement quantum, $M$ is the reference mass~\cite{giustino2007electron}, and $\epsilon_{mn}$ is the eigen-energy difference between states $m$ and $n$.

Eq.~\ref{eq:C_ij_ab_1_geo_s_main} clearly unveils the quantum geometry of IFM as the imaginary component of $\bm{G}^{s}_{mn}$, ${\rm Im}[\bm{G}^{s}_{mn}]$, termed the \textit{EPC Berry curvature}. It represents the Berry magnetic field in the hybrid Hilbert space and is directly associated with the inter-band matrix element of phonon-induced polarization via $\bm{p}^s_{mn} = -e \bar{Q}_s {\rm Im}[\bm{G}^{s}_{mn}]$~\cite{onoda2004topological,citro2023thouless,hu2024phononmediatednonlinearopticalresponses}. Therefore, the contribution of $C^s|_1$ to the phonon energy $\hbar \omega_s | _1$ can be evaluated as:
\begin{equation}\label{eq:hbaromegas_1}
    \begin{aligned}
        \hbar \omega_s | _1 &= \frac{1}{2} C^s|_1 \bar{Q}^2_s 
        = -\frac{1}{2} \int [d\bm{k}] \sum_{mn} f_m \bm{p}^s_{mn} \cdot \bm{\mathcal{E}}^s_{nm},
    \end{aligned}
\end{equation}
where $\bm{\mathcal{E}}^s_{nm} = \bar{Q}_s \bm{\gamma}^s_{mn} / e$ represents the matrix element of the phonon-induced electric field. Eq.~\ref{eq:hbaromegas_1} takes the typical form of the interaction energy between the electric dipole moment and the electric field, but in a quantum mechanical scheme, expressed in terms of inter-band matrix elements. This highlights the coherent interplay between valence and conduction states. A significant aspect is that the fully occupied subsystem, while provides no net contribution to the total polarization $\bm{P}^s_{\rm ele} = \int [d\bm{k}] \sum_{mn} f_m \bm{p}^s_{mn}$, can still contribute to the phonon energy through constructive interference with certain unoccupied states. Such coherent contributions are often significantly overlooked if one focuses solely on $\bm{P}^s_{\rm ele}$, or equivalently the Born effective charge, when analyzing lattice dynamics.


\textit{Geometric aspect of FE phase transition.} ---Based on the above investigation of EPC quantum geometry in IFM, we can further reveal the regulation mechanism of FE phase transition. The structural stability relies on positive $C^s = C^s|_1 + C^s|_{2,3} > 0$, which ensures real phonon frequencies via $\omega_s \propto \sqrt{C^s}$. Here, $C^s|_{2,3}$ are the eigenspace counterparts of the classical contributions $C^{ab}_{ij}|_{2,3}$ in Eq.~\ref{eq:raw_Cs_main}, which typically do not undergo sharp transitions~\cite{bersuker2013jahn, polinger2015pseudo, chen2022microscopic, SI_info} under symmetry-preserving regulations. In contrast, according to Eqs.~\ref{eq:C_ij_ab_1_geo_s_main} and~\ref{eq:hbaromegas_1}, the polarity reversal of the EPC Berry curvature, i.e., ${\rm Im}[\bm{G}^{s}_{mn}] \rightarrow -{\rm Im}[\bm{G}^{s}_{mn}]$, can result in a complete inversion of its contribution to $C^s|_1$, thereby potentially driving structural instability.

To illustrate this point, we begin with a general two-orbital electronic subsystem that is weakly entangled with other electronic states in the static crystal lattice. Its Hamiltonian takes the common form $\hat{H} = \epsilon_0 + \bm{\hat{\sigma}} \cdot \bm{h}$, where $\bm{\hat{\sigma}} = (\hat{\sigma}_x, \hat{\sigma}_y, \hat{\sigma}_z)$ are the Pauli matrices and $\bm{h} = (h_x, h_y, h_z)$ are the corresponding components. The eigen-energies are given by $\epsilon_{\pm} = \epsilon_0 \pm \epsilon_{\bm{k}}$, with $\epsilon_{\bm{k}} = |\bm{h}|$. The corresponding eigenstates, represented in the basis $(\psi_1, \psi_2)^{\bm{T}}$, can be projected onto the Bloch sphere as:
\begin{equation}\label{eq:psi_pm}
    |\psi_{+}\rangle =
    \begin{bmatrix}
        e^{-i\phi/2}\cos\left(\frac{\theta}{2}\right) \\
        e^{i\phi/2}\sin\left(\frac{\theta}{2}\right)
    \end{bmatrix}, 
    |\psi_{-}\rangle =
    \begin{bmatrix}
        e^{-i\phi/2}\sin\left(\frac{\theta}{2}\right) \\
        -e^{i\phi/2}\cos\left(\frac{\theta}{2}\right)
    \end{bmatrix},
\end{equation}
where $\theta$ and $\phi$ are the polar and azimuthal angles of the Bloch sphere, defined as $\cos{\theta} = h_z/\epsilon_{\bm{k}}$ and $\tan{\phi} = h_y/h_x$, respectively. Under the low-energy approximation, in the half-occupation case with Fermi level $E_f$ lies inside the band gap of this subsystem $|E_f - \epsilon_0| < \epsilon_{\bm{k}}$, $C^s|_1$ is primarily contributed by the internal transition between $|\psi_+\rangle$ and $|\psi_-\rangle$. In the full-occupation case with $E_f - \epsilon_0 > \epsilon_{\bm{k}}$, according to Eq.~\ref{eq:C_ij_ab_braket}, the non-vanishing contribution of $|\psi_{\pm}{\rangle}$ arises from their interaction with other unoccupied states $|n{\rangle}$. In both cases, as detailed in Sec.~IV~A~\cite{SI_info}, the EPC Berry curvature exhibits the following essential features: (i) The polarity (sign) can be reversed by inverting the orbital energy difference $h_z$ through ${\partial_s}\theta$ and ${{\nabla}_{\bm{k}}}\theta$; (ii) The strength tends to diverge at quasi-degenerate points with $|h_{x,y,z}| \to 0$, which, according to Eq.~\ref{eq:C_ij_ab_1_geo_s_main}, makes their contribution to $C^s|_1$ significantly dominant compared to other electronic states that do not have quasi-degeneracy. 

Consequently, if this subsystem incorporates quasi-degenerate points where $|h_{x,y,z}| \to 0$, any external regulation (e.g., strain, doping, or electric field) that induces band inversion by tuning $h_z$ to $-h_z$ can completely flip $C^s|_1$ to $-C^s|_1$. Therefore as aforementioned, since the classical contributions $C^s|_{2,3}$ typically show only trivial and moderate dependence on external regulation, the polarity reversal of $C^s|_1$ is likely to lead to the formation of a soft phonon mode as $C^s \to -C^s$, thereby causing structural instability. A dynamic Rice-Mele model coupled with in-plane optical phonon vibrations is presented in Sec.~IV~B~\cite{SI_info} to provide an intuitive understanding.

We term this type of electronic subsystem as FE-inducing states. While their detailed structure varies across different materials, several general characteristics can still be identified: (i) They must have direct coupling with phonon vibrations, enabling a significant response to IFM; (ii) They are often located at high-symmetry positions in both the Brillouin zone and real space, where (quasi) degeneracy is more likely to occur; (iii) They can correspond to fully occupied valence subsystems, allowing for band inversion without closing the global energy gap. This feature extends the applicability of our theory to large-gap semiconductors. In particular, according to (i) and (ii), the soft mode typically breaks inversion/mirror symmetry since it tends to lift the (quasi) degeneracy of FE-inducing states. As a result, the corresponding phase transition usually generates electric polarization with the same symmetry group representation, categorizing it as a FE type.

\textit{Application to BiOX.}---As a practical illustration, we apply our theory to BiOX (X = Cl, Br, I) monolayers, which adopt the representative fluorite structure~\cite{li2018oxygen,zhang2006study,cheng2014engineering,hussain2024recent} as shown in Fig.~S2(a,b). The DFT results in Fig.~\ref{fig:structure_frequency}(a) and Fig.~S2(c) indicate that when an in-plane biaxial tensile strain exceeds a critical threshold (approximately 2.6\% for BiOCl), an $E_u$ phonon mode (denoted by index $\parallel$) softens completely, signaling a strain-induced FE phase instability. To investigate the underlying mechanism, we begin by analyzing the band structure (Fig.~S3 and Fig.~S5) and the crystal orbital Hamiltonian population (COHP) (Fig.~S4), which reveal a three-band subsystem that correlates strongly with both the FE displacement and strain modulation. According to Fig.~S5~\cite{SI_info}, it is primarily composed of O's $p_{x,y,z}$ orbits. As shown in Fig.~\ref{fig:structure_frequency}(b), band inversion occurs near the critical strain, characterized by the change of $p_z$'s relative weight. 

To evaluate the impact of this band inversion, we construct a three-band tight-binding (TB) model, which, as detailed in Sec.~V~B~\cite{SI_info}, offers a great description to DFT results. Using the TB model, we reveal the sign reversal of the orbital energy difference \( m \) (schematicized in Fig.~\ref{fig:structure_frequency}(b)) and calculate the corresponding EPC Berry curvature, which exhibits polarity reversal near the critical strain, as shown in Fig.~\ref{fig:structure_frequency}(c). These results are consistent with our theoretical framework, confirming that the FE phase transition is driven by band inversion through EPC quantum geometry. Further validation is provided by calculating the strain dependence of the soft phonon energy using the TB model, which agrees well with the DFT results (Fig.~S2(a)). 

\begin{figure}[htp]
    \centering
    \includegraphics[width=0.48\textwidth]{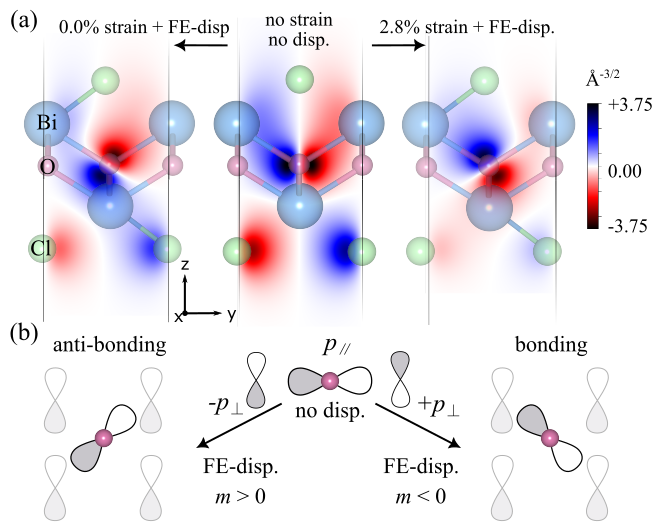}
    \caption{\label{fig:wavefunc_geometry}\textbf{Wavefunction interpretation of FE phase instability in BiOCl.} \textbf{(a)} DFT-calculated wavefunction of the $p_{xy}$-dominated branch in FE-inducing state, with the real part shown as the representative. In the absence of strain and FE displacement (referred to as FE-disp. in the figure), the wavefunction is primarily composed of O's $p_{xy}$ orbits with fully in-plane-polarized phase. With zero and 2.8\% strain (below and beyond the critical value), the FE displacement hybridizes $p_{xy}$ orbit with $+p_z$ and $-p_z$ orbitals, respectively. As illustrated in \textbf{(b)}, this difference leads to out-of-phase and in-phase alignments with neighboring Bi and Cl orbitals (shown with transparency), which results in anti-bonding and bonding response, respectively. }
\end{figure}

The polarity reversal of the EPC Berry curvature and its impact on FE instability can also be understood through the bonding response of wavefunctions with distinct phases. When the FE displacement $Q_{\parallel}$ introduces a perturbation $vQ_{\parallel}$ that is much smaller than the orbital energy difference $m$, the FE-inducing state can be approximated as~\cite{SI_info}: 
\begin{equation}
    \begin{aligned}
    |\psi_d{\rangle} &\ {\sim} \ i{\cdot}{\rm sign}(Q_{\parallel})|p_{\parallel}{\rangle} - \frac{|2vQ_{\parallel}|}{m}|p_{\perp}{\rangle},
    \end{aligned}  
\end{equation}
where $|p_{\parallel}{\rangle} = (|p_x{\rangle} + |p_y{\rangle})/\sqrt{2}$ and $|p_{\perp}{\rangle} = |p_z{\rangle}$. The sign of $m$ determines how the $|p_{\perp}{\rangle}$ orbit hybridizes with $|p_\parallel{\rangle}$. As shown in the DFT results and the schematic in Fig.~\ref{fig:wavefunc_geometry}, when $m > 0$ (or $m < 0$), the phase of $|\psi_d{\rangle}$ either mismatches (or matches) with the phase of the nearest-neighboring electrons, which results in an anti-bonding (or bonding) response to the ionic displacement, thus favoring (or disfavoring) the polar phase energetically.

\textit{Conclusions.}---In summary, this Letter reveals the critical role of the EPC Berry curvature in governing lattice dynamics and driving FE phase transitions. External regulations that induce band inversions can trigger FE phase transitions or other forms of structural instability by altering the polarity of the EPC Berry curvature. This mechanism is demonstrated using a general model and further verified by explaining the strain-regulated FE instability in the BiOCl monolayer. The theory is then extended to a wide range of materials, including BiOBr, BiOI, Bi$_2$O$_2$Se, BiCuOSe, Bi$_3$O$_4$Br, PbClF, and PbO, all of which exhibit strain-induced FE phase transitions accompanied by band inversions, as shown by DFT calculations in Sec.~VI~\cite{SI_info}. Since the theory arises from the singular behavior of degenerate electronic states, it provides a quantum geometric interpretation of the (pseudo) Jahn-Teller effect~\cite{bersuker2013jahn,bersuker2013pseudo,polinger2015pseudo} in crystalline solids. We therefore anticipate that this approach will have broad applicability to a wide range of materials exhibiting FE phase transitions and band crossings~\cite{liu2016strain,song2017first,PhysRevB.110.094408,chadov2010tunable}.


\textit{Acknowledgments.}---The authors acknowledge the insightful discussion with Dr. Wei Zhu in the School of Science, Westlake University. H.W. acknowledges the support from the NSFC under Grants Nos. 12304049 and 12474240, as well as the support provided by the Zhejiang Provincial Natural Science Foundation of China under grant number LDT23F04014F01. K. C. acknowledges the support from the Strategic Priority Research Program of the Chinese Academy of Sciences (Grants Nos. XDB28000000 and XDB0460000), the NSFC under Grants Nos. 92265203 and 12488101, and the Innovation Program for Quantum Science and Technology under Grant No. 2024ZD0300104. J.H. and Z.Z. contributed equally.

\bibliography{references}

\end{document}